\RequirePackage{fix-cm}
\documentclass[smallextended]{svjour3}       
\usepackage{natbib}

\smartqed  
\usepackage{graphicx}
\journalname{Software Testing, Verification and Reliability}
\usepackage{amsmath,amssymb,amsfonts}
\usepackage{algorithmic}
\usepackage{textcomp}
\usepackage{rotating}
\usepackage{multirow}
\usepackage{booktabs}
\usepackage{tabularx}
\usepackage{mdframed}
\usepackage{longtable}

\usepackage{listings}
\usepackage{xcolor}
\usepackage{framed}
\usepackage{caption}
\usepackage{subcaption}

\makeatletter

\usepackage{csquotes}

\usepackage{inconsolata}

\usepackage{color}

\definecolor{pblue}{rgb}{0.13,0.13,1}
\definecolor{pgreen}{rgb}{0,0.5,0}
\definecolor{pred}{rgb}{0.9,0,0}
\definecolor{pgrey}{rgb}{0.46,0.45,0.48}

\usepackage{listings}
\usepackage{hyperref}

\begin{document}

\newcommand{\rev}[1]{\textcolor{blue}{#1}}

\newcommand\setrow[1]{\gdef\rowmac{#1}#1\ignorespaces}
\newcommand\clearrow{\global\let\rowmac\relax}
\clearrow

\title{A new perspective on the competent programmer hypothesis through the reproduction of bugs with repeated mutations}
\titlerunning{A new perspective on the competent programmer hypothesis}

\author{
Zaheed Ahmed \and
Eike Stein \and
Steffen Herbold \and
Fabian Trautsch \and
Jens Grabowski}

\institute{
Zaheed Ahmed\\
Institute of Computer Science, University of Goettingen, Germany\\
University of Kotli Azad Jammu and Kashmir, Pakistan\\
\email{zaheed.ahmed@cs.uni-goettingen.de} (Corresponding Author)
\vspace{5pt}\\
Eike Stein\\Institute of Computer Science, University of Goettingen, Germany\\
\vspace{5pt}\\
Steffen Herbold\\Faculty of Computer Science and Mathematics, University of Passau, Germany\\
\email{steffen.herbold@uni-passau.de} 
\vspace{5pt}\\
Fabian Trautsch\\Institute of Computer Science, University of Goettingen, Germany\\
\vspace{5pt}\\
Jens Grabowski\\Institute of Computer Science, University of Goettingen, Germany\\
\email{grabowski@cs.uni-goettingen.de}
}

\date{Received: date / Accepted: date}

\maketitle

\begin{abstract}
The competent programmer hypothesis states that most programmers are competent enough to create correct or almost correct source code. Because this implies that bugs should usually manifest through small variations of the correct code, the competent programmer hypothesis is one of the fundamental assumptions of mutation testing. Unfortunately, it is still unclear if the competent programmer hypothesis holds and past research presents contradictory claims. Within this article, we provide a new perspective on the competent programmer hypothesis and its relation to mutation testing. We try to re-create real-world bugs through chains of mutations to understand if there is a direct link between mutation testing and bugs. The lengths of these paths help us to understand if the source code is really almost correct, or if large variations are required. Our results indicate that while the competent programmer hypothesis seems to be true, mutation testing is missing important operators to generate representative real-world bugs.

\end{abstract}

\section{Introduction}


Mutation testing is a concept for the evaluation of the quality of test suites through the systematic introduction of small variations into the source code, i.e., mutations~\citep{Papadakis.2019}. These variations result in many altered programs, called mutants. The rules by which program code is mutated are called mutation operators. The mutants that are generated by these operators should be similar to real software bugs. Some tests that previously succeeded should fail when executing mutated programs, otherwise they are considered inadequate in distinguishing correct programs from buggy ones~\citep{Offutt.1996}. If a mutant is detected by a test suite (by one or more tests failing after it is introduced) the mutant is \textit{killed}. If no test outcome changes, the introduced mutant survives. Mutation testing is based on two main assumptions~\citep{DeMillo.1978}.

\begin{itemize}
    \item The \textit{coupling effect} describes the correlation of test detection between different bugs. If a test is able to distinguish small changes introduced by mutation operators, then larger changes that are made up from multiple smaller changes will most likely also be detected by the same test~\citep{A.JeffersonOffutt.1992, R.A.Demillo.1990}. 
    \item The \textit{competent programmer hypothesis} claims that most programmers are competent enough to produce source code that is either correct or that only differs slightly from the correct code. Conversely, the small variations introduced by a mutation testing framework would then be able to mimic most real types of bugs. Therefore, if tests are not able to distinguish between the original code and mutants that mimic real bugs, then the test is not able to distinguish between correct and defective code either~\citep{Gopinath.2014, AllenT.Acree.1979, Hayes.2010, A.JeffersonOffutt.1992}.
\end{itemize}

There are multiple extensive and rigorous studies on the coupling effect \citep[e.g.,][]{Offutt.1989, TaiWah2001, andrews.2005, Just.2014, Laurent2022} that come to the same conclusion: the coupling effect exists and therefore, the capability of a test suite to detect mutants is correlated with the capability to detect real bugs. 

The evidence regarding the competent programmer hypothesis is not conclusive. For example, \cite{andrews.2005} found that mutants are similar to real bugs, while \cite{Gopinath.2014} found real bugs differed significantly from mutants. \cite{Papadakis.2019} note that there is a lack of new evidence regarding the competent programmer hypothesis in recent years. 

Due to the importance of the competent programmer hypothesis for the foundations of mutation testing, we believe that sound evidence is important to understand the underlying principles of mutation testing and define effective and efficient mutation testing strategies. Within this article, we provide new evidence regarding the competent programmer hypothesis and use a novel approach to study the relation between real-world bugs and mutation testing. Instead of directly comparing real-world bugs with mutants, we rather try to determine if we could reproduce bugs through the repeated application of mutation operators. Such repeated mutations are typically referred to as higher-order mutants~\citep{Papadakis.2019}. We apply graph-based path finding to the Abstract Syntax Trees (ASTs) of the clean and buggy problems. The paths themselves consist of chains of mutation operations. If we can find a chain of mutation operators that transforms the AST of a fixed bug back to the original bug, this means that there is a higher-order mutant that could exactly reproduce the bug. Otherwise, the bug cannot be reproduced by the available mutation operators. Hence, the existence of the path allows us to directly study the similarity of bugs with higher-order mutants. Consequently, the research question we try to answer within the article is as follows. 

\begin{description}
    \item[\textbf{RQ:}] Can real software bugs be recreated by higher-order mutants?
\end{description}

The answer to this question is directly related to the competent programmer hypothesis. If we can recreate the bugs, the small mistakes that are introduced by mutations are not only correlated, but representative for real-world bugs, which would support the hypothesis. Otherwise, we would have evidence against the competent programmer hypothesis, as not only single mutations would not be representative for real bugs, but also their combinations. 

The contributions of our study are the following. 

\begin{itemize}
    \item We found that while the competent programmer hypothesis seems to be true, commonly used mutation operators are not representative for real-world bugs and cannot be reliably used to reproduce these bugs. 
    \item We identified the addition of new method calls and blocks as the key aspect that mutation testing is missing to be more similar to real bugs. Learning application specific mutation operators could be a feasible solution to introduce such operators, as general operators for a programming language are likely not effective. 
\end{itemize}

The remainder of the article is structured as follows. We discuss the related work in Section~\ref{sec:relatedwork} followed by our approach in Section~\ref{sec:approach}. Section~\ref{sec:experiments} describes our experiments, including the data, mutation operators, measurements, and results. We discuss our results in Section~\ref{sec:discussion} and consider threats to the validity of our work in Section~\ref{sec:threats}. Finally, we conclude in Section~\ref{sec:conclusion}.

\section{Related Work}
\label{sec:relatedwork}

The focus of the discussion of the related work is on the competent programmer hypothesis. For a general discussion of the literature on mutation testing, we refer readers to the recent review by \cite{Papadakis.2019}. 

While the competent programmer hypothesis was studied in the past, there are still open questions. When \cite{DeMillo.1978} proposed mutation testing they recognized that mutation testing relies on the assumption that \enquote{Programmers have one great advantage that is almost never exploited: they create programs that are \textit{close} to being correct!}. However, \cite{DeMillo.1978} provided no proof that this assumption is true. Initial empirical works to provide evidence for the correctness of the competent programmer hypothesis were conducted by \cite{R.A.Demillo.1990} and \cite{Daran.1996}. In their studies, they manually evaluated sets of faults from single projects and came to different results. While \cite{R.A.Demillo.1990} found that about 20\% of bugs were simple, \cite{Daran.1996} found that about 85\% of bugs were simple and in line with the competent programmer hypothesis. Later, \cite{andrews.2005} came to the conclusion that real bugs are similar to mutants generated by mutation operators, but that mutants were harder to detect. However, \cite{Namin.2011} raised several concerns regarding the study by \cite{andrews.2005}, e.g., due to the choice of mutation operators and the selection of seeded faults for a single project. Therefore, while most early work seems to support the competent programmer hypothesis, it is unclear how the results generalize due to the limited scope of the studies and the problems with the validity that were raised. 

To the best of our knowledge, the only large scale analysis on the competent programmer hypothesis was conducted by \cite{Gopinath.2014}. They evaluated over 4000 open source projects in C, Java, Python and Haskell and analyzed the number of changed tokens by bug fixes tracked through the issue tracker on GitHub. They found that most real bugs differed significantly from the correct program version and concluded that \enquote{\dots our understanding of the competent programmer hypothesis, at least as suggested by typical mutation operators, may be incorrect} \citep{Gopinath.2014}. However, there are two potential issues with the validity of these results. First, just because something is marked as bug in an issue tracker, does not mean that this is really a bug. Research shows that about 40\% of bugs are mislabeled and actually requests for improvements or other changes~\citep{Herzig2013, Herbold2022}. Second, the analysis through tokens may overestimate the differences. A single mutation operator can modify multiple tokens. Additionally, tangled changes could lead to an additional overestimation of the difference, because there is a non-trivial amount of unrelated changes within bug fixing commits~\citep{Herzig2016, Herbold2022a}. Finally, we believe that \cite{Gopinath.2014} use a very strict interpretation of the competent programmer hypothesis to come to their conclusion. They assume that the competent programmer hypothesis requires single mutations, i.e., first-order mutants, to be sufficient to reproduce bugs. We disagree with such a strict reading of the hypothesis, because this would basically mean that competency and small variations are only achieved, if bugs can be fixed by touching a single line of code, usually without modifying the complete line. This would also mean that higher order mutants are not in line with the competent programmer hypothesis, which we also believe is too strict. 

We want to overcome the weaknesses of the study by \cite{Gopinath.2014} in two ways. First, we use a set of validated bug fixes as foundation for our analysis. This avoids noise due to mislabeled issues and tangling. In comparison to other work with validated data \citep{R.A.Demillo.1990, Daran.1996, andrews.2005}, our analysis is not limited to a single project. Second, we try not to measure the difference in tokens, but rather the order of mutation required to reproduce the bug. This way, we not only determine how many mutations are required, but also if the mutation operators we use are sufficient to reproduce bugs. The drawback of our work in comparison to \cite{Gopinath.2014} is that the scope is smaller, i.e., we consider only seventeen Java projects. Thus, our study should be seen as complementary to the work by \cite{Gopinath.2014}, with greater construct and internal validity at the cost of external validity. 

\section{Approach}
\label{sec:approach}

In this section, we present our approach to study the relation between mutation operators and real-world bugs. Our study explicitly does not cover if the mutation operators we use are ``good'' or ``bad'' in the sense of ``if these mutations are killed, we will likely also be able to find real bugs,'' as this would be a study of the coupling effect. Instead, the main question is whether it is possible to find a sequence of mutations that can recreate real bugs from corrected source code. Through this, we not only want to study if this is possible with a given set of mutation operators, but also which additional mutation operators may be useful to create mutations that mimic real bugs. Additionally, our approach should enable us to analyze the necessity for higher-order mutation operators or whether first-order mutation operators are sufficient. This means that we are interested in a shortest sequence of mutations that can recreate a bug. 

\begin{figure}
\centering
\includegraphics[width=\textwidth]{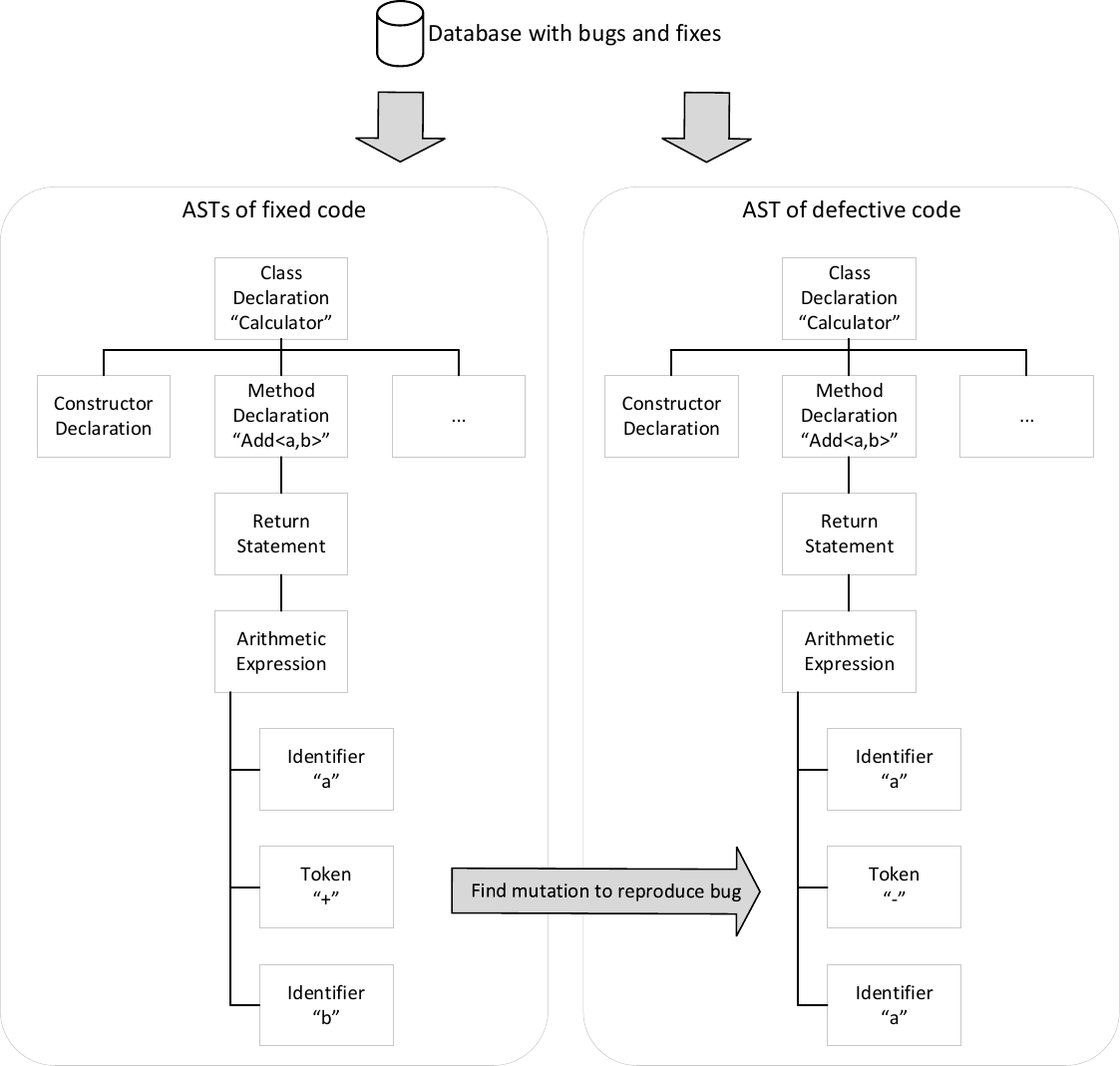}
\caption{General approach with an example AST.}
\label{fig:approach}
\end{figure}

To achieve this, we propose an algorithm that takes a pair of source code files and returns a sequence of mutation operators, together with the locations where they were applied and additional information about the values changed. Figure~\ref{fig:approach} summarizes our approach. Based on a database that contains bugs and the associated bug fixes, we create the ASTs of the fixed and defective files. Then, we use path finding to find a sequence of mutations that transforms the AST of the fixed file into the defective file. We now describe this approach in greater detail.

\subsection{AST Transformations}

Instead of trying to find a set of mutation operators that would mutate a file of source code from a fixed to a defective state directly, we first transform the source codes into ASTs. Applying mutation operators on the AST instead of the source code directly has several benefits. We avoid issues due to whitespaces or comments. Furthermore, with an AST parser in place, the implementation of mutation operators is a lot simpler, because ASTs are easier to reason about. In contrast to this, the comparison of source code directly can be challenging, because string matching is prone to noise \citep{MengyaZheng.2018}. Moreover, the usage of ASTs allows us to determine sequences of mutation operators that could reproduce bugs without actually mutating the source code. Instead, we have a graph transformation problem, where the mutation operators become modifications of the AST, e.g., the addition, deletion, replacement, or movement of AST nodes. Through the repeated application of mutation operators, we can achieve more complex transformations of the AST. 

We can formally describe this as follows. Let $S_{fix}$ be the corrected source file and $S_{bug}$ the source file that contains the real-world bug with $ast(S_{fix})$ and $ast(S_{bug})$ the corresponding ASTs. Moreover, we define a set of mutation operators $M = \{m_1, ..., m_n\}$ as functions $m_i: \mathcal{A} \to \mathcal{A}$ where $\mathcal{A}$ is the space of all possible abstract syntax trees and $i=1, ..., n$. A successful transformation of $S_{fix}$ into $S_{bug}$ is possible, if there is a finite sequence $p = (m^1, ..., m^k) \in M^k$ such that 

\begin{equation}
p(ast(S_{fix})) = (m^k \circ m^{k-1} \circ ... \circ m^1)(ast(S_{fix})) = ast(S_{bug}). 
\end{equation}

We also refer to a sequence of mutations as \textit{path}. Please note that we use $p$ to refer to the path both as a sequence and as the concatenation of the mutation operators. The length of the path $k$ is then the order of the mutation, i.e., if we need a path of $k$ mutations, we have a $k$-th order mutant. 

For example, consider a set of two mutation operators $M = \{m_1, m_2\}$. The first operator $m_1$ can replace the AST nodes \texttt{Token "+"} with the node \texttt{Token "-"}. The second operator $m_2$ can replace the wrong identifier in the last node \texttt{Identifier "a"} with \texttt{Identifier "b"}. The path $(m_1$, $m_2)$ can then mutate the AST in Figure~\ref{fig:approach} such that the bug is reproduced. Thus, there is a second-order mutant that can reproduce this bug. 

\subsection{Finding Mutation Sequences}

The existence of the path tells us if the error made by the developer that led to the bug can be reproduced by the artificially defined subset of errors that are the mutation operators. The length of the path indicates the number of simple errors required by the mutations model to reproduce real bugs, i.e., which order of mutations is required to generate mutants that closely resemble real bugs.

The naive approach would be to exhaustively search the space of possible paths, starting from the shortest path with a breadth first search. This means that we would first consider all mutation paths of length $k=1$, then of length $k=2$, etc. However, the size of the search space grows exponentially with the number of mutation operators. Additionally, mutation operators can usually not only be applied to a single AST node, but to many AST nodes. Even in the short example from Figure~\ref{fig:approach}, the replace identifier mutation could be applied to two nodes. Thus, such an exhaustive search is not feasible. 

Instead, we decided to use an A* algorithm \citep{Hart.1968} to search the space of possible mutations paths efficiently. The A* algorithm searches the possible paths with a \textit{best-first} approach. Paths are generated based on the costs of a path so far $g$ and an estimation of the remaining costs $h$ to reach the target. With suitable functions to estimate the costs $g$ and $h$, the A* algorithm is asymptotically optimal \citep{Dechter.1985}. 

To formulate our problem as a path search problem, we define a directed graph $G = (V, E)$ with nodes (vertices) $V \subset \mathcal{A}$. The set of edges $E \subset \mathcal{A}\times\mathcal{A}$ represent the mutations. For two nodes $ast_1, ast_2 \in V$ a directed edge $(ast_1, ast_2)$ exists, if and only if $m(ast_1)=ast_2$ for any mutator $m \in M$. We define the costs of a path as the number of mutations required, which is equivalent to the number of edges of the path in the graph problem, i.e., 

\begin{equation}
\label{eq:pathlength}
g(p) = k.
\end{equation}

The estimation of the remaining costs is more difficult, as this requires us to estimate how many more mutations are required to reach the target AST. The exact value can, by definition, only be known if we already know the minimal path. Thus, we need an indirect way for the estimation of these costs. In our case, we rely on GumTree \citep{Falleri2014}, a tool that can determine a path of transformations between ASTs to perform AST differencing. In comparison to our approach, GumTree can freely manipulate the ASTs, whereas we are limited to the modifications supported by the set of mutation operators $M$. Due to this, the paths of GumTree may be completely different from our mutation paths. Regardless, the length of the GumTree path is a good measure for the AST similarity, as a shorter GumTree path means that fewer modifications are required to finish a mutation path. Thus, we can estimate the remaining effort for a path $p$ as 
\begin{equation}
h(p) = \textit{ast\_diff}(p(ast(S_{fix})), ast(S_{bug}))    
\end{equation}
where $\textit{ast\_diff}: \mathcal{A} \times \mathcal{A} \to \mathbb{N}$ is the number of edit operations determined by the AST differencing algorithm of GumTree. 

\subsection{Partial Reproduction of Bugs}

Because the mutations of the AST are limited by the expressiveness of the mutation operators in $M$, it is likely that not all bugs $S_{bug}$ can be reproduced through a mutation path $p$ starting from $ast(S_{fix})$. However, how much of the difference between $ast(S_{fix})$ and $ast(S_{bug})$ can be reduced by a path $p$ is also important information for our study. We use the AST differencing of GumTree for this to measure the \textit{progress} of the path $p$ towards a complete reproduction of a bug as

\begin{equation}
\label{eq:progress}
\textit{progress}(p) = 1-\frac{\textit{h(p)}}{\textit{ast\_diff}(ast(S_{fix}), ast(S_{bug}))}.
\end{equation}

Since the A* algorithm minimizes $h(p)$, we can still use the algorithm to find a mutation path with the best progress towards reproducing the bug, even if the algorithm does not converge and find a complete path. We modified the algorithm to store partial paths in a cache, once the path search fails to find a successor that is a better solution. 

\section{Experiments}
\label{sec:experiments}

We conducted an experiment based on  Defects4J data \citep{Just.2014} to get insights on the competent programmer hypothesis. The nature of our experiment is exploratory. The reason for this is that the prior work on the competent programmer hypothesis is inconclusive and does not allow for the derivation of a clear hypothesis that we could confirm. Instead, our goal is to provide additional evidence regarding the competent programmer hypothesis that could be confirmed by future work. In the following, we describe the data, mutation operators, evaluation measures, methodology of the experiments, and results. Our results and implementation are available online.\footnote{https://github.com/sherbold/replication-kit-chp-study}

\subsection{Data}

The foundation of our experiments is the Defects4J dataset first published by \cite{Just.2014}. Table~\ref{tab:analysisprojects} gives an overview of the data of 17 projects provided in the currently available version of Defects4J (v2.0.0)\footnote{https://github.com/rjust/defects4j}. Overall, we used 835 bugs in our experiments. Unfortunately, we failed to execute our approach on all available bug-fix pairs due to instrumentation issues, e.g., because GumTree failed to find a path between two ASTs as the change was too large. Due to this, we had to exclude 33 bug-fix pairs from the dataset for our experiments. 

Since our approach is limited to the modification of a single AST, we cannot directly apply it for the reproduction of bugs that affect multiple files. There are three possible solutions:
\begin{enumerate}
    \item exclude all bugs that affect multiple files;
    \item consider a bug as reproduced if and only if we determine that the AST of all affected files can be transformed; or
    \item consider all files individually and consequently treat bugs that affect multiple files as multiple bugs.
\end{enumerate}
The first approach would reduce our data and also skew the data to possibly simpler bugs, further affecting the validity of the results. The second approach has the advantage, that we would still consider bugs as the atomic units of our analysis. However, the results would be harder to interpret and possibly hide valuable information. For example, if a bug affected three files and two can be fully reproduced, but the third file would not be reproduced, this valuable information would be hidden. This is the advantage of the third approach, i.e., we see for each file change if it can be reproduced. The drawback is that our analysis unit effective becomes partial bugs. However, in most cases exactly one file was changed, so there is only a small amount of noise. We only have 25\% more file changes than bugs. Consequently, we decided for the third approach and report the results for each file changed as part of a bug fix. Overall, we have 1044 file changes. 

\begin{table}
\centering
\begin{tabular}{lccc}
\textbf{Project ID} & \textbf{Bugs Total} & \textbf{Bug-Fix Pairs Total} & \textbf{Bug-Fix Pairs Included}\\
\hline\hline
Chart & 26 & 28 & 25 \\
Cli & 39 & 51 & 51 \\
Closure & 174 & 225 & 220 \\
Codec & 18 & 24 & 24 \\
Collections & 4 & 4 & 4 \\
Compress & 47 & 58 & 58 \\
Csv & 16 & 17 & 17 \\
Gson & 18 & 21 & 21 \\
JacksonCore & 26 & 35 & 30 \\
JacksonDatabind & 112 & 160 & 155 \\
JacksonXml & 6 & 6 & 5 \\
Jsoup & 93 & 125 & 124 \\
JxPath & 22 & 35 & 35 \\
Lang & 64 & 64 & 57 \\
Math & 106 & 119 & 114 \\
Mockito & 38 & 46 & 46 \\
Time & 26 & 26 &  25 \\
\hline
\textbf{Total} & \textbf{835} & \textbf{1044} & \textbf{1011}\\
\end{tabular}
\caption{Number of analyzed pairs of bugs and fixes per project}
\label{tab:analysisprojects}
\end{table}

\subsection{Mutation Operators}

The choice of mutation operators directly affects our analysis. A large set of mutation operators is more expressive and could reproduce a larger number of bugs. However, our path finding algorithm A* has a worst case exponential memory complexity $O(b^d)$, where $b$ is the \textit{branching factor}, i.e., the average number of successors of a step in the path search and $d$ is the length of the path. Thus, we must find a middle ground: a sufficiently large set of representative mutation operators for meaningful results, while also not using any operator possible to bound the exponential nature of the A* algorithm by controlling the branching factor. 

We decided to use a tool-driven approach for the selection of a suitable set of mutation operators. Our rationale was that mutation operators used by tools should be designed with the competent programmer hypothesis in mind and, therefore, be representative for small variations that should be expected based on the competent programmer hypothesis for a given technology, in our case Java. 

\begin{description}
\item[MuJava] is a well-known mutation testing frameworks for Java \citep{Ma.2005} and has many operators from a variety of classes. The main focus of MuJava lies on mutation operators for object-oriented programming and the extensibility with more mutations. However, MuJava also provides a set of method-level operators. These operators are, in principle, similar to the operators of other frameworks, in the sense that they modify or delete the existing arithmetic operations or logical statements. 
\item[Major] is another mutation testing framework with a strong focus on extensibility and customizability. Instead of applying the mutations on a source code level, Major can integrate itself into the compiled byte code at compilation time~\citep{major}. One big difference between Major and other mutation testing frameworks is the ability to define and configure mutations through the domain specific language Mml \citep{majordocumentation}. However, Major does not provide a guidance regarding a subset of suitable mutation operators that are recommended for practical use. Due to the exponential nature of our approach, it is vital that we have a reasonably small set of mutations. While we could determine a subset from Major on our own, we rather want to re-use an existing set of mutations without selections on our part that could lead to problems with the reliability of our research. 
\item[Pitest] (or PIT) was originally intended to extend JUnit tests to allow them to be run in parallel. After that problem was solved, the authors decided to use their enhancement to apply mutations during the test execution and, thereby, developed their project to a fully functional mutation testing framework \citep{pitFAQ}. Due to the easy integration into the build process, Pitest is one of the drivers of the practical relevance of mutation testing, highlighted by frequent mentions in developer forums and blogs. Pitest has four sets of mutation operators, with a default set of eleven operators \citep{pitoperators}.
\end{description}
We also evaluated other Java mutation testing frameworks (e.g., Javalance \citep{Schuler.2009}, Jumple \citep{ReelTwo.15.05.2015}, and Jester \citep{IvanMoore.2001}) but discarded them as options because they are no longer actively developed.

Based on our assessment of the tools, we decided to use the state-of-the-art default and optional (i.e., old defaults, stronger and all) mutation operators sets of Pitest \footnote{http://pitest.org/quickstart/mutators/} as foundation for our work, which we summarize in Table~\ref{tab:pitestoperatorsdescription}. These mutation operators are widely used and very expressive for logical changes that are conducted within methods, i.e., mutations that do not change the interface of methods and, consequently, classes. We also note that these operators or similar operators are also available in Major and MuJava. In the following, we refer to the default set of mutation operators as $M_{default}$ and the combination of $M_{default}$ and optional operators as $M_{optional}$. We defined five additional mutation operators as given in Table~\ref{tab:extendedoperatorsdescription}, that should help to fix some short-comings of $M_{default}$ and $M_{optional}$, especially with respect to finding mutations that require specific values of literals or where references should be replaced. These are also simple mistakes that are in line with the competent programmer hypothesis (i.e., failing to rename variables after copy \& paste) that would otherwise not be covered. We refer to the combination of $M_{default}$, $M_{optional}$ and the five new operators as $M_{extended}$. We note that there is sometimes an overlap between the operators in the set, e.g., that the operator ``Relaxed Empty Return'' could always be used instead of the simpler ``Empty Returns''. Due to this, we bias the A* heuristic to first use operators from the default set, then the optional set, and only then from the extended set. 

We actively decided against additional operators, especially those that modify the interface of methods and classes. Our rationale was that such mistakes would not be in line with the competent programmer hypothesis, as changes that affect the design of a software cannot reasonably be characterized as small differences to the correct code. Our decision to not use such operators was also a factor in favor of Pitest over MuJava as foundation for our work. 

\begin{table}
\centering
\begin{tabular}{lp{6.5cm}}
\textbf{PIT Mutation Operator} & \textbf{Description} \\
\hline\hline
Conditionals Boundary & Replaces relational operators such as $>$, $>=$, $<$ and $<=$ with a different relational operator. \\
Increments & Replaces increments with decrements and vise versa. \\
Invert Negatives & Inverts the value of floating point and integer values, either applied to a variable or directly to hard coded values. \\
Math & Replaces a mathematical operator ($+$, $*$, $\%$, $|$, \dots) with another mathematical operator. The plus operator in string concatenations ("A"+"B") is excluded because it is not considered to be a mathematical operator. \\
Negate Conditionals & Similar to the conditionals boundary operator, this operator replaces relation operators, by inverting them. $==$ becomes $!=$ and so forth. \\
Void Method Calls & Deletes calls to methods that do not return any values. \\
Empty Returns & Replaces the value in a return statement with the default value for that type. For example strings become empty strings, integers and floating point numbers become 0.  \\
False Returns & Replaces Boolean return statements to always return false. \\
True Returns & Replaces Boolean return statements to always return true. \\
Null Returns & Replaces reference type return statements to always return null. \\
Primitive Returns & Replaces numeric return statements to always return 0. \\
\hline
Return Values & In the new default group, this operator has been replaced by another set of return mutators i.e. empty, false, true, null, and primitive returns. Depending on the return type of a method, it mutates the return value.\\
Remove Conditionals & Ensures that guarded statements never/always execute by mutating the equality checks (e.g., $==$, $!=$) and order checks (e.g., $<$, $>=$, $>$, $<=$).\\
Inline Constant & Depending on the type of a non-final variable, it replaces literal values with another value.\\
Constructor Calls & Places null values as a substitute for constructor calls.\\
Non Void Method Calls & Replaces non-void method calls with the Java default values according to the method return type.\\
Remove Increments & Removes increment ($++$) and decrement ($--$) operators from the local variables.\\
Negation & Replaces any numeric variable with its negation.\\
Arithmetic Operator Replacement & Similar to the math operator, but it replaces one arithmetic operator (e.g., $+$) with all other operators from this set (e.g., $-$, $*$, $/$, $\%$). \\
Arithmetic Operator Deletion & Removes binary arithmetic operator along with one of its operands.\\
Constant Replacement & Similar to the inline constant operator, it substitutes a constant c with 1, 0, $-$1, $-$c,  c$+$1, and c$-$1.\\
Bitwise Operator & Removes bitwise operator (e.g., $\&$, $|$) along with one of its operands. It also replaces $\&$ with $|$ and vice versa.\\
Relational Operator Replacement & Similar to the negate conditionals operator but it replaces one relational operator (e.g., $<$) with all other operators from this set (e.g., $<=$, $>$, $>=$, $==$, $!=$).\\
Unary Operator Insertion & Inserts an increment ($++$) or decrement ($--$) operator to a local variable.\\
\hline
\end{tabular}
\caption{Pitest mutation operators we used. The description of default and optional sets of Pitest \citep{pitoperators} is separated by the line.}
\label{tab:pitestoperatorsdescription}
\end{table}

\begin{table}
\centering
\begin{tabular}{lp{6.5cm}}
\textbf{Extended Mutation Operator} & \textbf{Description} \\
\hline\hline
Method Calls & The default set of Pitest can only mutate code to remove void method calls. We added an additional operator that can remove any method call, as otherwise all bugs where non-void method calls are removed would not be reproducible. This approach is similar to an experimental mutation operator of Pitest for non-void method calls.\\
Relaxed Empty Returns & The empty returns operator uses a fixed default value as replacement. While such a replacement may be a reasonable mutant, it is unlikely that such a mutant matches a real-world bug exactly because the replacement value may be different. We modified the empty returns operator to allow any possible replacement value to be able to better reproduce bugs. \\
Relaxed Inline Constants & Same as relaxed empty returns, but for the inlining of constants. \\
Relaxed Return Values & Similar to relaxed empty returns. Additionally, we not only allow all possible values of the same literal type are possible, but also the replacements of object references. Otherwise coding mistakes where the wrong object was returned could not be reproduced.\\
Rename & A generic operator that can replace valid identifier names with other valid identifier names, e.g., the name of a method with another method. We added this operator, because otherwise all bugs where a method call is replaced could not be reproduced.\\
\hline
\end{tabular}
\caption{Extended mutation operators we used for additional experimentation.}
\label{tab:extendedoperatorsdescription}
\end{table}

\subsection{Measurements}

Let $B$ be the set of bugs included in our study with $b = (S_{fix}, S_{bug}) \in B$ as pair of fixed and defective source code. We define our performance metrics based on three subsets of the bugs $R$, $P$, and $U$ that are defined as

\begin{equation*}
R = \{(S_{fix}, S_{bug}) \in B: \exists~p \in M^*~\text{with}~p(ast(S_{fix}))=ast(S_{bug})\}
\end{equation*}
\begin{align*}
P = \{(S_{fix}, S_{bug}) \in B: \exists~p \in M^*~\text{with}~&\textit{ast\_diff}(p(ast(S_{fix})), ast(S_{bug})) \\<~&\textit{ast\_diff}(ast(S_{fix}), ast(S_{bug}))\} \setminus R
\end{align*}
\begin{equation*}
U = B \setminus (R \cup P)
\end{equation*}
where $M^*$ is the the set of all possible mutation paths. 

These sets define the bugs that we can fully reproduce ($R$), that we can partially reproduce ($P$), and that we are unable to reproduce ($U$) using mutation operators. The size of these sets is our measure to explore the competent programmer hypothesis: if we can reproduce or at least partially reproduce bugs, this indicates support for the competent programmer hypothesis. We use the progress of partial paths as defined in Equation~\eqref{eq:progress} to gain further insights into how well the partially reproduced bugs support competent programmer hypothesis. Bugs that cannot be reproduced at all are contradictory to the competent programmer hypothesis. 

The second measure we use is the length of the mutation paths, i.e., the order of the resulting mutants for bugs that we can reproduce. Short mutation paths would provide strong support for the competent programmer hypothesis for these bugs. Longer paths indicate that the bugs are the sum of many small mistakes, which means that the bug was actually not just almost correct, which would not be in line with the competent programmer hypothesis. 

\subsection{Methodology}

Our experiment methodology is straight forward and consists of three phases. In all phases, we apply our approach for the reproduction of bugs through mutations described in Section~\ref{sec:approach} for the Defects4J data. In the first phase of the experiments, we use $M_{default}$ to determine how many bugs from Defects4J we can reproduce, in the second phase we use $M_{optional}$. Thus, the first phase evaluates the relationship between a standard set of mutations that is already used in practice and the support of the competent programmer hypothesis. The second level analyses how the Pitest optional operators influence the recreation of bugs ratio when used along with the $M_{default}$ set. The third phase evaluates how our extension of the allowed mutations $M_{extended}$ impacts the results. We used insights from the first two phases to extend the set of mutation operators. Through this, we mitigate a potential impact on our results due to a small and restricted set of mutation operators. In all phases, we measure the relation between the competent programmer hypothesis and the set of mutations through the measurements defined above. 

\subsection{Results}

\begin{figure}
\centering
\includegraphics[width=\textwidth]{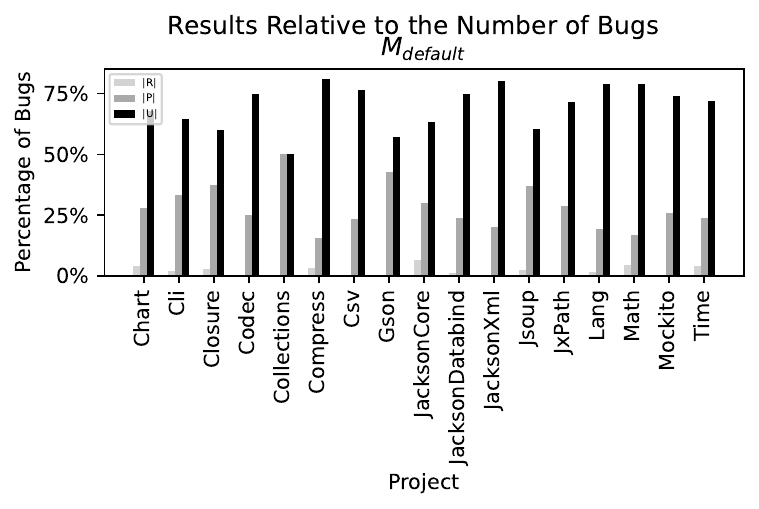}
\includegraphics[width=\textwidth]{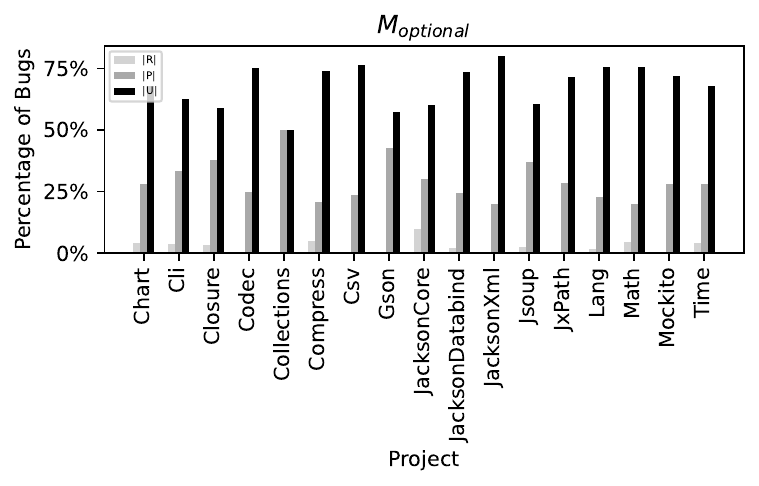}
\includegraphics[width=\textwidth]{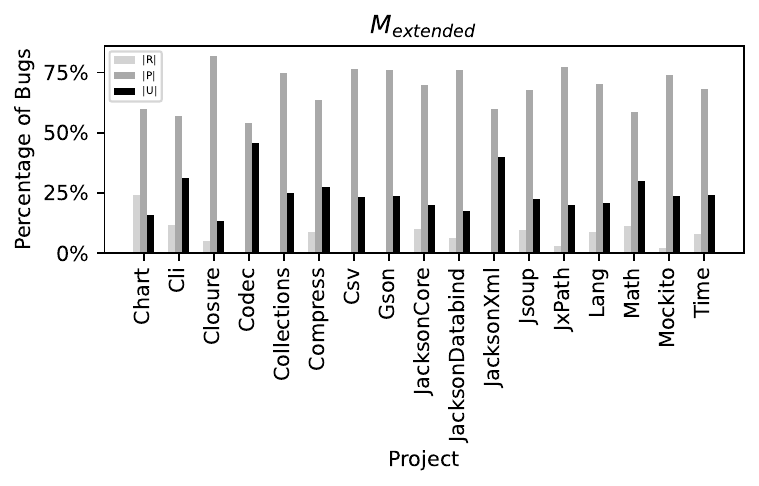}
\caption{Percentages of bugs that were reproduced (R), partially reproduced (P), and not reproduced (U).}
\label{fig:bugs-relative}
\end{figure}

\begin{figure}
\centering
\includegraphics[width=\textwidth]{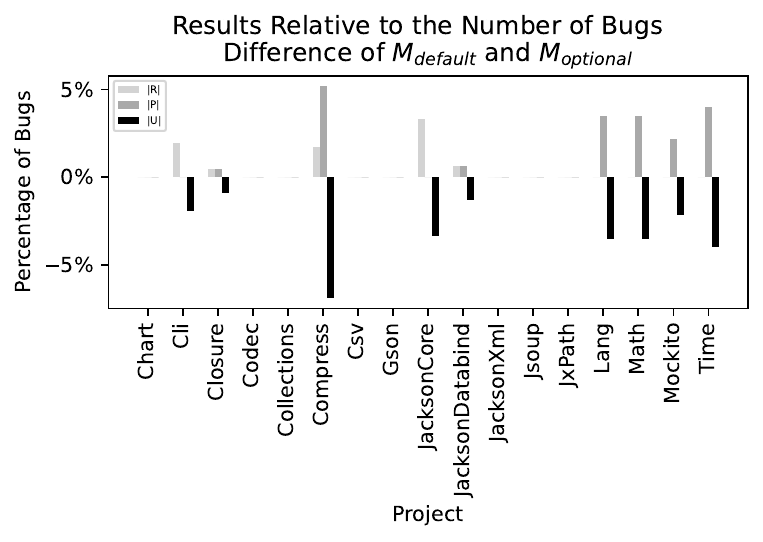}
\includegraphics[width=\textwidth]{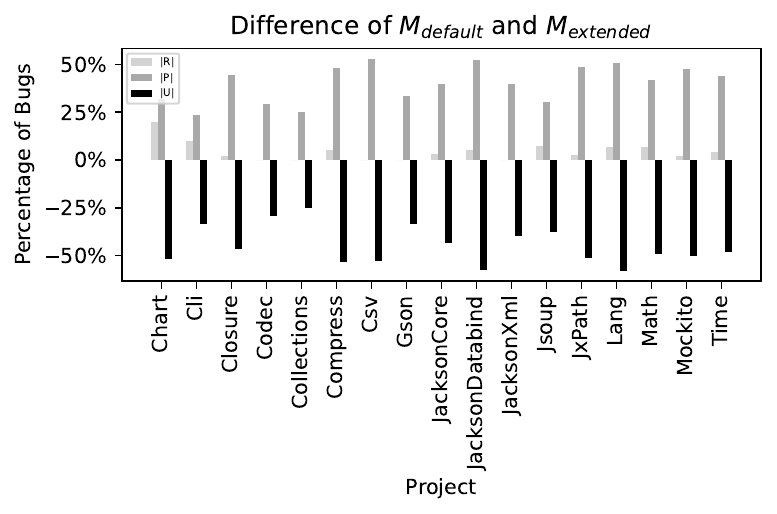}
\includegraphics[width=\textwidth]{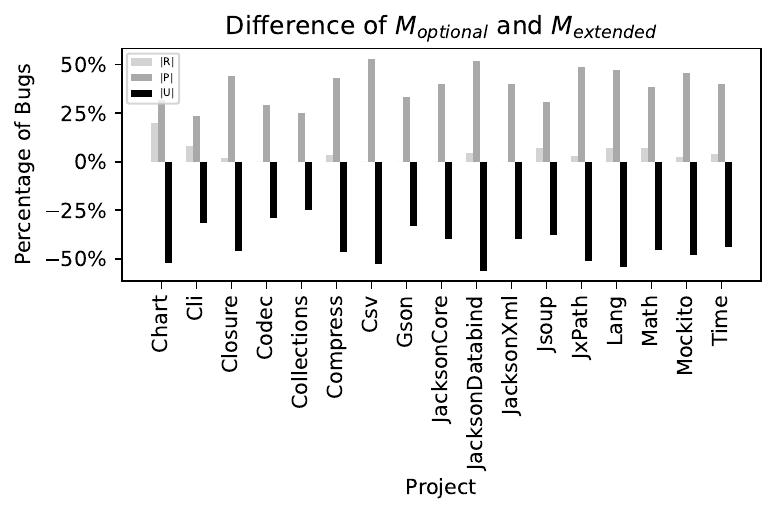}
\caption{Difference: Percentages of bugs that were reproduced (R), partially reproduced (P), and not reproduced (U).}
\label{fig:bugs-relative-difference}
\end{figure}

Table~\ref{tbl:results} shows the absolute number of our reproduction of bugs. Overall, we could fully reproduce 24 bugs with the $M_{default}$ operators, 29 bugs with the $M_{optional}$ operators and 75 bugs with the $M_{extended}$ operators. Moreover, 287 bugs could at least be partially reproduced with the $M_{default}$ operators, 300 with the $M_{optional}$  and 717 with the $M_{extended}$ operators. We could not reproduce 700 bugs with the $M_{default}$, 682 bugs with the $M_{optional}$ and 219 bugs with the $M_{extended}$ operators. Figure~\ref{fig:bugs-relative} provides a relative view on the data, i.e., the percentage of bugs per project that could reproduced, partially reproduced, or not reproduced at all. Figure~\ref{fig:bugs-relative-difference} shows the difference among the percentage of bugs reproduced by using $M_{default}$, $M_{optional}$ and $M_{extended}$ sets.  The data is relatively stable and shows that between 50\% to 81\% percent of the bugs could not be reproduced at all with the $M_{default}$ mutation operators. This changes with the $M_{extended}$ operators, that allows for the partial reproduction of 54\% to 82\% of the bugs instead.  

\begin{figure}
\centering
\includegraphics[width=\textwidth]{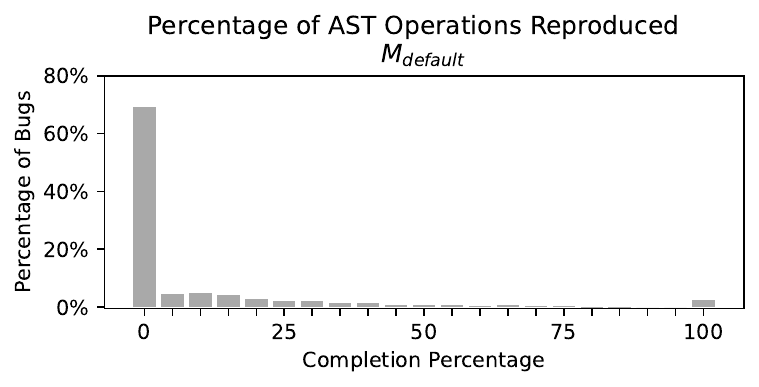}
\includegraphics[width=\textwidth]{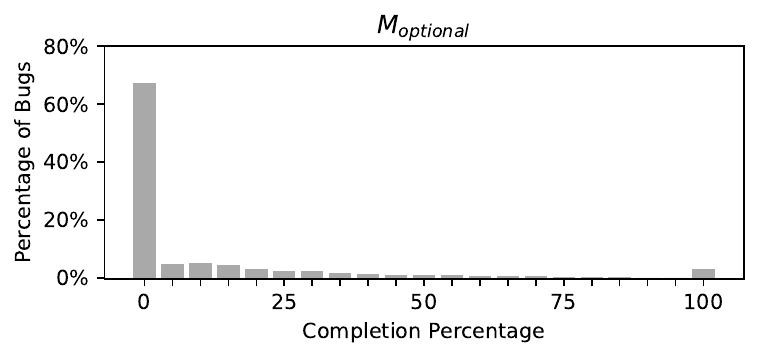}
\includegraphics[width=\textwidth]{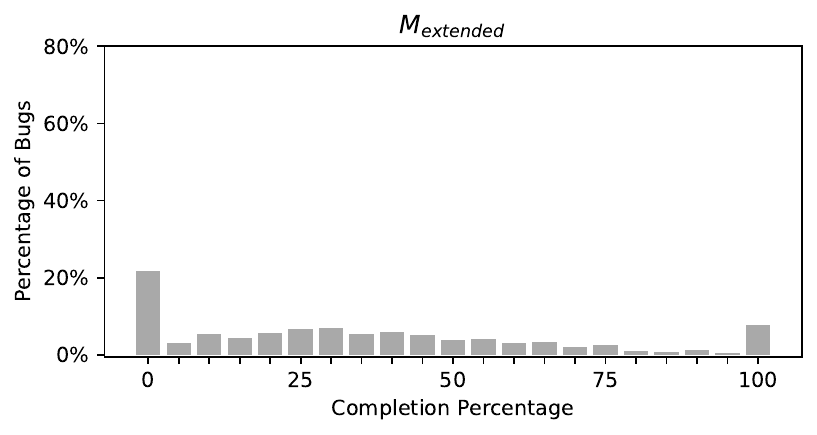}
\caption{Percentage of required AST changes that were reproduced through mutations.}
\label{fig:percentages}
\end{figure}

Figure~\ref{fig:percentages} allows us a closer look at the partial reproduction of bugs and shows which percentage of the required AST we were able to reproduce. With all the group of operators, i.e. $M_{default}$, $M_{optional}$ and $M_{extended}$, the results are similar, with the difference that the number of bugs reproduced by $M_{extended} > M_{optional} > M_{default}$. For bugs that can partially reproduce, we observe similar distributions: there are few cases where less than 10\% are partially reproduced, the partial reproduction peaks at about 30\% of the bugs reproduced, and then steadily declines for large percentages until there are only very few partial reproductions above 75\%. The peak and the decline are a bit faster for the $M_{default}$ and $M_{optional}$ operators, which is also in line with the smaller amounts of bugs that we could partially reproduce. 

\begin{figure}
\centering
\includegraphics[width=\textwidth]{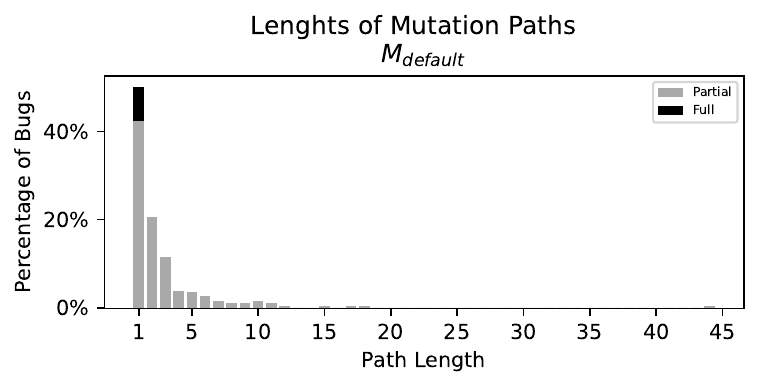}
\includegraphics[width=\textwidth]{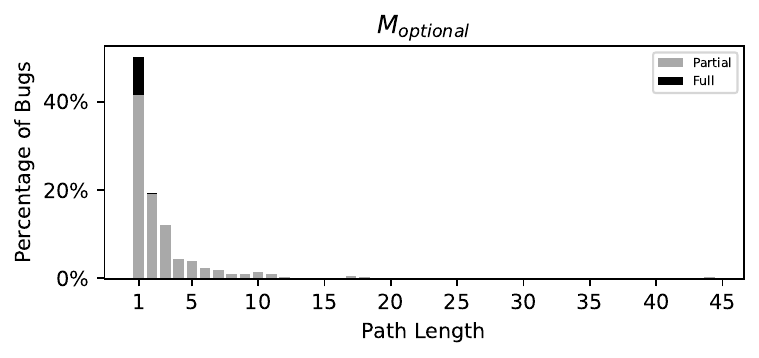}
\includegraphics[width=\textwidth]{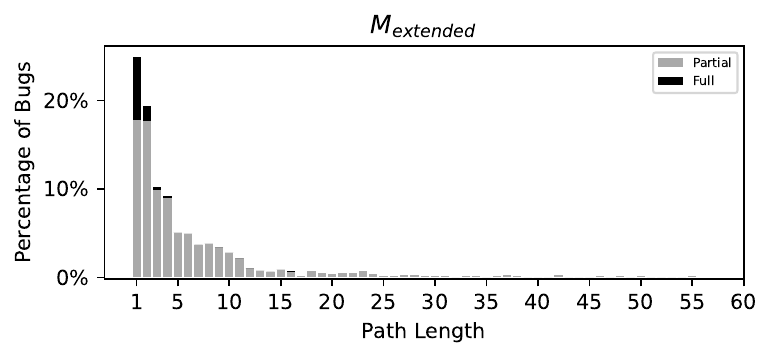}
\caption{Lengths of the mutation paths we found to reproduce bugs. Maximum path length for $M_{default}$ and $M_{optional}$ is 44 while for $M_{extended}$ its 55. }
\label{fig:lengths}
\end{figure}

Figure~\ref{fig:lengths} shows the length of the mutation paths, i.e., the order of mutations we determined for the reproduction of bugs. With the $M_{default}$ and $M_{optional}$ operators, we observe that full reproduction was only achieved through first-order mutants except one second-order mutant for the $M_{optional}$. This is different for the $M_{extended}$ operators, where we also see full reproductions with up to fourth-order mutants with one outlier, i.e., a sixteenth-order mutant. For the partial reproduction, we observe that there are many cases where longer mutation paths are considered. The longest mutation path that is found consists of 55 mutations. However, we also observe a strong decay of the lengths of the mutation paths, i.e., while we observe about the same number of first-order and second-order mutants for partial reproduction with $M_{extended}$, the number of higher order mutants steeply declines for higher mutation orders. With the $M_{default}$ and $M_{optional}$ sets, we observe this steep decline already for second-order mutants. 

Finally, Table~\ref{tbl:ops-used} reports which mutation operators were selected for the mutation paths. The operators ``Increments'', ``Empty Returns'', ``Remove Conditionals'', ``Inline Constant'', ``Constructor Calls'', ``Relaxed Emtpy Returns'', ``Relaxed Inline Constants'', and ``Relaxed Return Values'' are never used. The most effective operators for the reproduction of bugs are the ``Void Method Calls'' and ``Method Calls'' operator, i.e., the removal of method calls. The ``Rename'' operator was also often used to switch identifier names. We further find that ``Relational Operator Replacement'' seems to be important, but is only frequently used with the $M_{extended}$ set. We observe the same for the ``Negate Conditional'', ``Non Void Method Calls'' and ``Arithmetic Operator Deletion'' operators. ``Negation'' operator is only used with the $M_{extended}$ set. ``Conditional Boundary'', ``Math'', ``False Returns'' and ``Null Returns'' are equally used in all mutation operators sets. Application of ``Constant Replacement'' and ``Bitwise Operator'' operators is noticed similar in both $M_{optional}$ and $M_{extended}$ sets. ``Invert Negatives'', ``True Returns'' and ``Remove Increments'' are valuable, but less important than the other operators. The ``Primitive Returns'', ``Arithmetic Operator Replacement'', ``Return Values'' and ``Unary Operator Insertion'' operators are also helpful, but not for many bugs. Relaxing the return operator did not make a difference, as this only merged the results for ``False Returns'' and ``True Returns'' into a single operator.

\begin{table}
\centering
\begin{tabular}{l | ccc | ccc | ccc}
& \multicolumn{3}{c|}{$M_{default}$} & \multicolumn{3}{c|}{$M_{optional}$}& \multicolumn{3}{c}{$M_{extended}$} \\
\textbf{Project ID} & $|R|$ & $|P|$ & $|U|$ & $|R|$ & $|P|$ & $|U|$ & $|R|$ & $|P|$ & $|U|$\\
\hline\hline
Chart & 1 & 7 & 17 & 1 & 7 & 17 & 6 & 15 & 4 \\
Cli & 1 & 17 & 33 & 2 & 17 & 32 & 6 & 29 & 16 \\
Closure & 6 & 82 & 132 & 7 & 83 & 130 & 11 & 180 & 29 \\
Codec & 0 & 6 & 18 & 0 & 6 & 18 & 0 & 13 & 11 \\
Collections & 0 & 2 & 2 & 0 & 2 & 2 & 0 & 3 & 1 \\
Compress & 2 & 9 & 47 & 3 & 12 & 43 & 5 & 37 & 16 \\
Csv & 0 & 4 & 13 & 0 & 4 & 13 & 0 & 13 & 4 \\
Gson & 0 & 9 & 12 & 0 & 9 & 12 & 0 & 16 & 5 \\
JacksonCore & 2 & 9 & 19 & 3 & 9 & 18 & 3 & 21 & 6 \\
JacksonDatabind & 2 & 37 & 116 & 3 & 38 & 114 & 10 & 118 & 27 \\
JacksonXml & 0 & 1 & 4 & 0 & 1 & 4 & 0 & 3 & 2 \\
Jsoup & 3 & 46 & 75 & 3 & 46 & 75 & 12 & 84 & 28 \\
JxPath & 0 & 10 & 25 & 0 & 10 & 25 & 1 & 27 & 7 \\
Lang & 1 & 11 & 45 & 1 & 13 & 43 & 5 & 40 & 12 \\
Math & 5 & 19 & 90 & 5 & 23 & 86 & 13 & 67 & 34 \\
Mockito & 0 & 12 & 34 & 0 & 13 & 33 & 1 & 34 & 11 \\
Time & 1 & 6 & 18 & 1 & 7 & 17 & 2 & 17 & 6 \\
\hline
\textbf{Total} & \textbf{24} & \textbf{287} & \textbf{700} & \textbf{29} & \textbf{300} & \textbf{682} & \textbf{75} & \textbf{717} & \textbf{219}\\
\end{tabular}
\caption{Absolute numbers of bugs that we recreated through mutations. $|R|$ are fully reproduced, $|P|$ partially, and $|U|$ not reproduced at all.}
\label{tbl:results}
\end{table}

\begin{table}
\centering
\begin{tabular}{lrrr}
\textbf{Mutation Operator} & $M_{default}$ & $M_{optional}$ & $M_{extended}$ \\
\hline\hline
Conditionals Boundary & 10 & 10 & 10\\
Increments & 0 & 0 & 0\\
Invert Negatives & 1 & 1 & 3\\
Math & 4 & 4 & 4\\
Negate Conditionals & 9 & 9 & 14\\
Void Method Calls & 786 & 790 & 831\\
Empty Returns & 0 & 0 & 0\\
False Returns & 3 & 3 & 3\\
True Returns & 2 & 2 & 2\\
Null Returns & 3 & 3 & 3\\
Primitive Returns & 0 & 0 & 1\\
\hline
Return Values & - & 1 & 1\\
Remove Conditionals  & - & 0 & 0\\
Inline Constant  & - & 0 & 0\\
Constructor Calls  & - & 0 & 0\\
Non Void Method Calls  & - & 11 & 12\\
Remove Increments  & - & 2 & 2\\
Negation  & - & 0 & 5\\
Arithmetic Operator Replacement  & - & 0 & 1\\
Arithmetic Operator Deletion  & - & 8 & 9\\
Constant Replacement  & - & 4 & 4\\
Bitwise Operator  & - & 6 & 6\\
Relational Operator Replacement  & - & 11 & 16\\
Unary Operator Insertion  & - & 1 & 1\\
\hline
Method Calls & - & - & 2091\\
Relaxed Empty Returns & - & - & 0 \\
Relaxed Inline Constants & - & - & 0 \\
Relaxed Return Values & - & - & 0 \\
Rename & - & - & 1383\\
\end{tabular}
\caption{Number of times mutation operators were used in all phases of the experiment. A hyphen sign indicates that the operator was not available during that part of the experiment.}
\label{tbl:ops-used}
\end{table}

\section{Discussion}
\label{sec:discussion}

The general message of our results regarding the research question is clear: we fail to recreate most real-world bugs through mutations, regardless of the order of the mutants. However, we can at least partially reproduce a large proportion of real-world bugs. Within this section, we dive deeper into our results and identify reasons why this was not possible, discuss what this means for the competent programmer hypothesis, and the implications of these results on the future of mutation testing. 

\subsection{Reasons for Failed Bug Reproductions}

The main reason for failed reproductions is that the mutation operators are not in line with how software is often modified as part of bug fixes. One aspect is that bug fixes are often related to changes of method calls. That more flexible deletion of method calls is valuable is already part of our results through the relaxed ``Method Calls'' operator. The ``Rename'' operator also covers cases in which a method was replaced with a different locally available method with the same signature. However, this still does not account for the addition of new method calls. Thus, all bugs that contained a method call that was not part of the fixed source code, could not be reproduced. We note that this also means that our decision to use the Pitest operators~\cite{Coles.2016} for this study does not have a strong impact on the results, as these bugs could also not be reproduced by the operators of any of the other mutation testing frameworks we are aware of. 

This problem can be generalized as the key reason for failed reproductions: whenever there is a change to an external dependency that is outside of the scope of a method that is modified as part of a bug fix, mutation testing is likely not able to reproduce such bugs. Unfortunately, the design of generic mutation operators that could solve this problem in a meaningful way is very hard, maybe even impossible. Consider what this means: mutation operators should be able to pick a suitable (!) candidate from all (!) possible method calls that could be inserted at any (!) possible locations. The number of possibilities is for all practical purposes infinite for non-trivial applications. Since the number of suitable candidates is likely very small, the chance of randomly selecting a good method call to insert is basically zero. However, this is how normal mutation operators work: they define a logic what should be mutated and then insert this mutation in the source code. Bounding this problem can be considered as the opposite of program repair, which indicates that learning common mistakes within an application may be a solution for this. This implies that the solution to this problem could be an application of a specific set of mutations that is learned from prior bugs. Interestingly, this seems to be exactly the approach suggested by \cite{Beller2021} in a recent paper on the adoption of mutation testing in practice. The first example of a learned mutation that \cite{Beller2021} present is a complex replacement of a method call.

A second pattern we found that would be hard to define with mutation operators is the addition or removal of conditional blocks, e.g., if statements with null checks. The inability by the mutations to add complete blocks hinders the reproduction of such bugs. Similar to the addition of method calls, there is a huge amount of possibilities, where new blocks could be inserted randomly. Moreover, such operators would also have to decide which statements should become part of the new block, while also generating meaningful conditions. Thus, random mutation operators are, again, likely not possible, while learning operators could also solve this problem. 

Thus, while we were initially surprised that the number of bugs we could fully reproduce is relatively low, a deeper inspection of the problem shows that this is not surprising and that fixed set of mutation operators are not likely to overcome this problem. Consequently, our analysis also showed that the inability is not due to the complexity of the bugs. 

\subsection{Competent Programmer Hypothesis}

Our evidence for the competent programmer hypothesis is not straightforward and the actual implications are a matter of interpretation both of our findings and the meaning of the competent programmer hypothesis. 

\begin{figure}
\centering
\includegraphics[width=0.5\textwidth]{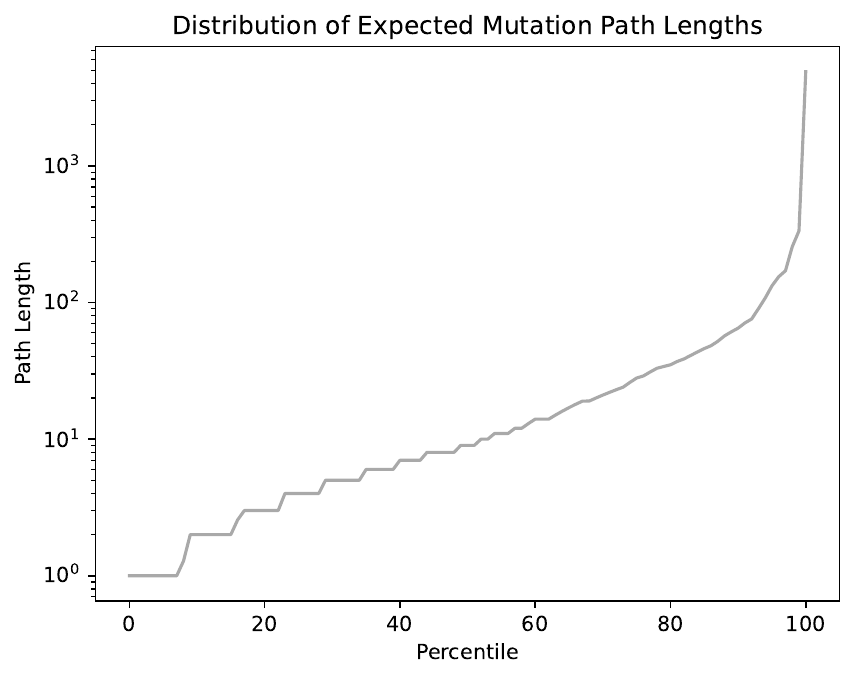}
\caption{Expected lengths of mutation paths required for the reproduction of bugs.}
\label{fig:lengths-expected}
\end{figure}

On the one hand, we found that many bugs can not be reproduced using mutation operators. However, as we discussed in the previous section, in many cases the root cause for this was not the complexity of the bug but rather that the mutation operators were not sufficiently powerful. On the other hand, we found when we could fully reproduce bugs, the mutation paths where relatively short. The partial reproductions also contained longer mutation paths, and mostly contained between 5\% and 75\% of the required mutations. We can estimate the order of mutations required for all bugs, where we have at least partial mutations for the path $p$ as
\begin{equation}
\label{eq:estimate}
\textit{estimate}(p) = \frac{\textit{g(p)}}{\textit{progress(p)}}.
\end{equation}
Where $g(p)$ and the $progress(p)$ are defined in Equation~\eqref{eq:pathlength} and Equation~\eqref{eq:progress} respectively. This provides the expected mutation path lengths, assuming that the same ratio of the required mutations are missing. We then proceed and add the data for fully reproduced bugs. Figure~\ref{fig:lengths-expected} shows the result. We observe that the sequence lengths seem to be exponentially growing with the percentiles. In other words, we have an exponential decay for the lengths of mutation paths: 40\% of the bugs should be reproducible by 7 or less mutations, 50\% of bugs should be reproducible by 9 or less mutations, 60\% of the bugs should be reproducible by 14 or less mutations and 70\% of bugs should be reproducible by 21 or less mutations. 

Thus, while we find 30\% of non-trivial bugs that require more than 21 mutations to reproduce, we still believe that our results support the competent programmer hypothesis. Most bugs can be replicated with only few mutations, which is in line with the competent programmer hypothesis. There is also gray area of bugs that require more, but not a lot of mutations. We do not want to judge whether these are still small variations or larger deviations. However, the competent programmer hypothesis does not state that mutation testing should be able to mimic \textit{all} real bugs, but rather only \textit{most} real bugs. Thus, we do not believe that the more complex bugs we found contradict the hypothesis, instead we postulate that these are the ``few'' bugs that are not among the ``most'' mistakes. However, same as we found that standard mutation operators do not cover all important cases, we also find that first-order mutations are too simple to represent real-world bugs. 

We note that our discussion is based on our interpretation of the competent programmer hypothesis. If we were to assume a rigid stance regarding the competent programmer hypothesis, similar to \cite{Gopinath.2014} our conclusion would be strongly against the competent programmer hypothesis. This shows a vital research question that should be addressed by future studies: what do developers interpret as ``small variation'' that are in line with the competent programmer hypothesis. Without such work, we can only provide evidence regarding the data, but not yet achieved a universally accepted interpretation of this evidence. 

\subsection{Implications}

While we support the competent programmer hypothesis, our results are rather on the side that the bugs introduced by currently used mutation operators are not similar to the majority of bugs. However, this is not because the bugs are complex, but rather because the mutation operators lack the expressiveness to cover the relevant AST modifications. 

We note that our results should not be seen as implication that the mutation operators, e.g., of Pitest, are bad. However, that they work and can be used to improve test suites can, to our mind, not be explained using the competent programmer hypothesis, as the bugs these operators mimic are not realistic. Instead, we can only speculate why current mutation testing is effective. One explanation could be that the coupling effect is sufficiently effective on its own. Another aspect could be that even unrealistic mutations are sufficient to uncover a lack of assertions in tests. Further research on the foundations of mutation testing is required to better understand why this works, e.g., by analyzing how test suites are improved as a consequence of mutation testing. 

The second, possibly more important, implication of our results is that research into the targeted generation of higher-order mutants could be valuable, as these seem to better mimic real-world bugs. Such mutations should especially cover how logic may change through additional calls of methods, which may be particularly hard to solve, due to the huge amount of possibilities. The challenge that must be solved is to identify new method calls such that
\begin{itemize}
\item the mutants are not live mutants because the additional method calls have no effect; and
\item the mutants are not trivial to kill, e.g., because they crash the application. 
\end{itemize}

Recent research by \cite{Beller2021} shows the promise of using real-world bugs to learn mutation operators, instead of specifying them manually through the addition. Such learned mutations generate effectively higher order mutants, as they provide more complex modifications of the AST. Since the patterns could even include how new methods are selected for mutation, we believe that such approaches may lead to a major advancement of mutation testing, both with respect to the effectiveness, as well as the adoption. 

Another interesting result of our study is that certain mutations were never used for the reproduction of bugs. While this does not automatically mean that such mutations cannot help, e.g., to improve test suites, this is certainly an indicator that not all mutation operators mimic real-world bugs that occur regularly. This is another indication that learning mutations could lead to more effective mutation testing. 

\section{Threats to Validity}
\label{sec:threats}

There are several threats to the validity of our work, which we report following the classification by \cite{Wohlin2012}.

\subsection{Construct Validity}

The core of our construct is the selection of mutation operators. An unsuitable set of mutation operators could alter our results towards finding fewer or longer mutation paths and, thereby, bias our results against support for the competent programmer hypothesis. We countered this threat through the extension of our set of mutation operators through additional operators that resolve restrictions and make the approach more general. Our data shows that this was effective, as we were much better able to reproduce bugs with the extended set of mutation operators. Since our results are sufficient to estimate the expected order of mutations to fully reproduce the bugs in our sample, and the data also shows the limitations of current mutation testing operators, we do not believe that the usage of other mutation operators would substantially alter our results. There could also be a threat to our results because we first use the default operators, then the optional operators and then the extended operators. However, this should not limit the capability to find mutation paths, but rather only lead to using simpler operators first.

\subsection{Internal Validity}

Our conclusion in support of the competent programmer hypothesis is, in part, based on extrapolation of our results, i.e., the combination of our findings regarding the observed path lengths and the completion percentage. While we made the reasonable assumption that a similar number of mutations is required to cover the remaining segment of AST modifications, this is not necessarily true. Instead, the remaining part could be more complex mutations that affect more AST nodes at once, therefore requiring fewer mutations. Regardless, while this means we may overestimate the order of mutations required to reproduce real bugs, this would not impact our findings: there would still be a non-trivial number of bugs that require a large number of mutations and most bugs would still be reproducible with relatively few mutations.

\subsection{External Validity}

Our experimentation is limited to data from Defects4J, which is not an unbiased sample of bugs in general, or even for Java software. Therefore, our results may not translate to other settings, e.g., other Java projects or projects written with other programming languages. However, the generalization of our results would only be affected, if the bugs from Defects4J are particularly simple or particularly difficult. If the bugs are simpler then bugs on average, this would mean that our conclusion in support of the competent programmer hypothesis may be wrong, as we would underestimate the number of mutations required. If the bugs are more difficult than bugs on average, our findings with respect to the competent programmer hypothesis would hold, but our extrapolation that learning mutation operators can overcome difficulties may not hold, as simple pre-defined operators may be sufficient. While we cannot mitigate this threat, we are also not aware of any research on the Defects4J data that indicates that these bugs are particularly simple or difficult, nor did we see any indications regarding this in our study. 

The size of our sample of bugs is another threat to the generalizability of our results. However, the trends we observe in our data are very clear and our sample size is with 835 unique bugs not small, i.e., we do not see any indication why they should change with a larger sample from the same population, e.g., more bugs sampled the same way as Defects4J.

\section{Conclusion}
\label{sec:conclusion}

Within this article, we consider the competent programmer hypothesis through the lens of our ability to reproduce bugs through mutation operators. This gives us a different perspective on the competent programmer hypothesis than prior work that, e.g., only considered how many tokens are changed by bugs or manually compared mutations with bugs. We re-framed the problem of transforming a correct into a buggy AST as a path search problem, in which each step of a path is a mutation. This allows us to not only evaluate if mutation operators are directly related to bugs, but also the order of the mutations required to fully reproduce bugs. We found that our data supports the competent programmer hypothesis. However, we were often not able to fully reproduce bugs, because the limited expressiveness of mutation operators. We found that especially the addition of new blocks and method calls is a large difference between real-world bugs and mutation operators that can only delete or modify AST nodes. Thus, while our results indicate that the competent programmer hypothesis is true, mutation operators are often not in line with the slight differences of correct code introduced by developers. 

In the future, we plan to investigate if automatically learned mutation operators are better suited to reproduce real-world bugs and, thereby, demonstrate that mutation testing is not only based on the competent programmer hypothesis, but actually fully in line with the real-world bugs. These investigations will help us to further understand why mutation testing works with the goal to at some point not just demonstrate correlation between mutations and real-world bugs, but actually find causal links between mutations and real-world bugs that can be exploited to improve the effectiveness and efficiency of mutation testing. 


\section*{Declarations}

The authors have no competing interests to declare that are relevant to the content of this article.

\bibliography{./literature}

\end{document}